\documentstyle[11pt,aaspp4]{article}
\input{psfig.sty}

\eqsecnum
\def\beq{\begin{equation}}
\def\eeq{\end{equation}}
\def\ref{\reference}
\def\about{$\sim$}
\def\erg/cm2sec{ergs~cm$^{-2}$~s$^{-1}$}
\def\ergcm2{ergs~cm$^{-2}$}
\def\arcmin{$\,^\prime$~}
\def\la{\hbox{\rlap{\raise.3ex\hbox{$<$}}\lower.8ex\hbox{$\sim$}\ }}
\def\ga{\hbox{\rlap{\raise.3ex\hbox{$>$}}\lower.8ex\hbox{$\sim$}\ }}
\def\deg{$^{\circ}$~}
\begin{document}

\title{Fast X-ray Transients and Gamma-ray Bursts: 
Constraints on Beaming}

\author{Jonathan E. Grindlay}
\affil{Harvard-Smithsonian Center for Astrophysics, 60 Garden St., 
Cambridge, MA 02138}

\begin{abstract} 
The detection of x-ray afterglows of gamma-ray burst (GRB) sources 
has enabled the optical discovery that GRBs  
are at cosmological distances, which confirms that the enormous luminosities 
required must produce a relativistically expanding fireball. It is 
not yet clear if this expansion is quasi-spherical or beamed, or if 
the beaming differs between the prompt GRB and afterglow, 
although the differences for both source emission and population models 
are profound.  We show that relative beaming 
(prompt vs. afterglow) may be constrained  
by analysis of the ``fast transient" x-ray sources catalogued
by Pye and McHardy (1983) from the {\it Ariel V} survey as 
well as several other (e.g. HEAO-A1/A2) surveys. After removing 
contamination from likely stellar transients, and scaling from 
x-ray/$\gamma$-ray fluence ratios recently derived from Ginga 
and BeppoSAX,  approximately half of the  sources were possible 
prompt emission or afterglows 
from GRBs with $\gamma$-ray fluence of \ga1.2 $\times$ 10$^{-5}$ \ergcm2. 
Since the rate of these candidate afterglow events, 0.15 day$^{-1}$, is 
consistent with the BATSE logN-logS distribution, 
no difference in prompt vs. afterglow beaming is required. 
However more sensitive x-ray monitors or 
hard x-ray imaging surveys, with better temporal and spatial resolution, could 
provide  strong constraints on GRBs and beaming.

\end{abstract}


\section{Introduction}

The recent discoveries of x-ray (Costa et al 1997) afterglow emission 
from gamma-ray burst (GRB) sources provided the long-awaited 
breakthrough needed for settling the distance scale. The fading 
soft/medium x-ray sources detected by BeppoSAX for several 
bursts were bright enough to be imaged and well positioned (\about1\arcmin) 
for several days after a GRB, enabling the first followup optical detection 
(Van Paradijs et al 1997) of a GRB. Subsequent measurement of 
the absorption redshift (z = 0.83) for GRB970508 by Metzger et al (1997) 
has confirmed the cosmological distance scale (for at least some GRBs) and 
the implied enormous energy release (\ga10$^{51}$ ergs) in a 
radiation-dominated  fireball, as hypothesized by 
Meszaros and Rees (1993, and references therein) and subsequently 
developed in many papers (e.g. Meszaros, Rees and Wijers 1998, 
hereafter MRW). The fireball must expand 
relativistically, with likely Lorentz factor $\Gamma$ \ga 10$^{2-3}$. 
A key question is whether the required relativistic expansion is 
quasi-isotropic or beamed, most likely into an opening angle 
$\sim1/\Gamma$.  The prompt GRB could appear similar (cf. MRW) 
in either case, although beam-edge effects could also be observable (cf. 
Panaitescu, Meszaros and Rees 1998), but 
of course the source energetics and  required numbers would be 
modified by factors of  $\Gamma^{-2}$ and $\Gamma^2$, respectively.

Whereas the prompt GRB emission is probably due to the 
initial expansion and deceleration of the fireball (and may also 
be governed by internal shocks), 
the subsequent afterglow emission 
likely results from the later deceleration of the expanding shock 
in the ambient ISM. Given the entrainment of the fireball 
and its deceleration by the ISM in the later phases,  
the afterglow is likely to be more 
isotropic than the prompt GRB emission 
(but see Rhoads (1997) and Dar (1998) 
for effects of afterglow beaming), 
and detection of GRBs by just their afterglow emission  
can test  {\it differential beaming}:  if a significantly larger rate of afterglow 
events is detected than expected (from log N - log S)  
for prompt GRBs of same total  fluence threshold, then 
afterglow beaming must indeed be less than prompt beaming.  
In principle afterglows could be 
detected by either sensitive optical/IR or x-ray searches. The 
present GRB optical counterparts suggest that  
optical/IR searches would need to be sensitive to \ga18-20th mag 
objects, fading to perhaps \ga23 mag over just \about3-10d. Thus 
deep supernova searches offer perhaps 
the best hope for detecting optical afterglows without an 
accompanying GRB trigger, but since these are restricted to 
very small (e.g. typically 10\arcmin) fields, the rate will be negligible 
unless indeed the prompt GRB emission is strongly and 
differentially beamed. [Indeed after the initial submission of 
this paper, and on the basis of the likely optical identification 
of GRB980425 with SN1998bw, Wang and Wheeler (1998) have suggested
that at least some GRBs are associated with SNIb/c's and that strong (differential) 
beaming is implied.]

In contrast, the soft-medium energy 
x-ray afterglows are much brighter and thus 
easier to detect amidst the multitude of background variable 
sources. The soft (0.2-2 keV) or medium (2-10 keV) energy 
x-ray afterglow fluxes are fading from an initial x-ray 
brightness that is apparently 0.05-0.25 as bright  
as the prompt $\gamma$-ray (50-300 keV; or BATSE trigger range) 
energy flux in the GRB itself (cf. Strohmayer et al 1998, 
Frontera et al 1998). That is, x-ray afterglows fade from 2-10 keV fluxes 
approximately as bright as the Crab ($\sim2 \times 10^{-8}$ \erg/cm2sec), 
or nearly the brightest source in the sky. In contrast,  
the optical transients appear to be \ga$10^3 \times$ fainter and so 
fade from a relatively faint (but still unsure; and subject 
to reddening) \ga18th mag. 

In this paper we utilize published catalogues of so-called ``fast 
transient"  x-ray (2-10 keV) sources to constrain the rate of 
GRBs detected by (just) their x-ray emission -- from 
both prompt x-ray emission and afterglow. Fast transients   
are \about2-10 keV x-ray sources detected on just 1-2 satellite orbits 
(i.e. \about1.5-3 h duration, typically) by virtually all the early 
generation of x-ray satellites (non-imaging) with sensitivities 
\about5 mCrab = 1 $\times 10^{-10}$ \erg/cm2sec (2-10 keV). 
Our primary 
data is from the {\it Ariel V} survey published by 
Pye and McHardy (1983), but we also consider the 
independent HEAO-1 surveys which incorporated the A1 
(Ambruster and Wood 1986) and A2 (Connors, Serlemitsos and 
Swank 1986) data.

\section{The {\it Ariel V} Fast Transients: GRBs ?}

Pye and McHardy (1983; hereafter PM) 
reported the cumulative results of 
the survey for fast transient x-ray sources which were detected 
(2-18 keV) only in only 1 to a few orbits (100min) 
with the Sky Survey Instrument 
(SSI) instrument on the UK x-ray satellite {\it Ariel V}. Their 
catalogue of 27 Ariel Transient (AT) sources provided a 
more comprehensive tabulation than the several previous 
reports (cf. references in PM) of similarly fast transients 
detected  with ANS, SAS-3 and HEAO-1.  The sources 
were required to be above a 5.5$\sigma$ threshold flux of 
8 SSI counts s$^{-1} \sim$ 20 mCrab (2-10 keV) 
= 4 $\times 10^{-10}$ \erg/cm2sec for their integrated 
flux over 1 orbit (which, at \about100 min,  was the minimum integration 
time scale) and to not correspond obviously to known persistent 
x-ray sources detected by {\it Ariel V} and catalogued 
in the 3A catalogue (Warwick et al 1981). Source locations 
were typically only measured to \about5-10 deg$^2$, given the 
0.75\deg $\times$ 10\deg collimator, which was inclined at 
65\deg to the plane scanned at the 6 s rotation period of the satellite.  

The AT catalogue contains 27 sources (including one source 
likely detected twice) with locations approximately 
isotropic on the sky and a flux distribution obeying a logN-logS 
relation N(\ga S) = K (S/S${_o}$)$^{-\alpha}$ 
where K = 65 $\pm30$ sources y$^{-1}$, $\alpha = 0.8 \pm0.8$, and S$_o$ 
= 8 SSI counts s$^{-1}$. These 27 sources 
were detected as significant excesses over 1 (or  at most a few; 
see below) orbits from some 20,000 satellite orbits, the usable 
total from the \about30,000 orbits 
accumulated over the full 5.5 year 
(October 1974 - March 1980) mission lifetime. 
The resulting full-sky rate of AT sources derived 
by PM is R$_{AT} \sim$ 0.3 day$^{-1}$ for sources detected with 
2-10 keV x-ray flux above the 20 mCrab threshold, integrated 
over 1 orbit. Since peak fluxes for the AT sources are not 
known (given the 100 min time resolution), we convert S$_o$ 
first to a 2-10 keV x-ray {\it fluence} threshold for AT events: 
F$_{X,min}$ = F$_{AT,min} 
\sim$ 6000 S$_o$ =  2.4 $\times 10^{-6}$ \ergcm2. 
\medskip

\noindent
{\it Prompt GRBs ?}

We first consider whether the AT events could be due to 
x-ray (2-10 keV) prompt emission from GRBs. In fact, 
PM noted that two of their 27 catalogued sources were 
coincident in time with GRBs: sources AT0318+220 = 
GRB751102 and AT0710+012 = GRB760128. Not only 
are these events temporally coincident, but for the second 
event the \about1\deg spatial coincidence is compelling (Cline et al 1979). 
The first event had no available position data reported 
(Klebesadel et al 1982). We note below  
that the long duration x-ray detections of each 
(3 and 2 satellite orbits, respectively) 
now provide strong evidence for x-ray afterglow detection of these 
historical GRBs.  
 
In order to compare the AT events in general to current BATSE 
studies of  GRBs, we need to convert 
the x-ray fluence to a likely $\gamma$-ray fluence in the 50-300 keV 
band used for BATSE triggers. This depends, of course, on the 
still uncertain x-ray and $\gamma$-ray spectra spectral shapes of 
GRBs.  BeppoSAX measures of the x-ray (2-10 keV) 
vs. $\gamma$-ray (40-700 keV) fluence 
of GRBs (e.g. Frontera et al 1998) now provides 
a followup to the historical sample of \about120 GRBs 
detected with the combined x-ray (2-25 keV) 
and $\gamma$-ray (15-400 keV) GRB detector 
(GBD; cf. Murakami et al 1989) which operated on the  
{\it Ginga} satellite from March 1987 until October 1991. 
Strohmayer et al (1998) have analyzed the sample of 
22 bright GRBs detected with the {\it Ginga} GBD and derive an 
average ratio of x-ray/$\gamma$-ray fluences 
r$_{X\gamma,Ginga}$ = 0.24. This is very comparable 
to the value quoted by Frontera et al (1998) for GRB970228, 
r$_{X\gamma,SAX} = 0.2\pm0.05$ although the 
bands (both x-ray and $\gamma$-ray) are different. 
Given the typical GRB spectral shape of a broken 
power law and exponential cutoff (e.g. the Band et al 1993 
model, also used by Strohmayer et al),  
the BeppoSAX bands are in fact close to the conversion 
we need for AT/BATSE fluences. Thus we adopt 
r$_{X\gamma,SAX}$  to estimate 
the corresponding 50-300 keV $\gamma$-ray fluence threshold for 
AT events (assuming they are prompt emission from GRBs): 
F$_{\gamma,min}$ = F$_{X,min}$/ r$_{X\gamma,SAX}   
 \sim 1.2 \times 10^{-5}$ \ergcm2. 

Thus if the AT sources are {\it prompt} x-ray emission 
from GRBs, they correspond to relatively bright bursts, 
with fluence threshold F$_{\gamma,min}$ \ga 30$\times$ 
that of BATSE (the relative sensitivity is uncertain due 
to the complications of trigger criteria). We 
consider below the detailed composition 
of the AT sources, but if they are all prompt x-ray 
counterparts of  GRBs, then their rate may be compared 
with the logN - logS rate for GRB fluence above 
F$_{\gamma,min} \sim 1.2 \times 10^{-5}$ \ergcm2. 
The equivalent BATSE detection rate for 
GRBs with fluence \ga F$_{\gamma,min}$  
then provides a crude upper limit for the prompt GRBs 
contained in the AT survey;  any  
significant AT-GRB rate above this would 
imply detection of GRBs by their afterglows alone. 
Using the corrected logN-logS for fluence presented by 
Bloom, Fenimore and in't Zand (1996) for the 
BATSE 3B catalogue, and normalization provided 
by Bloom (private communication, 1998), we obtain 
the expected rate 
R(F\ga F$_{\gamma,min}$) $\sim$ 0.23 day$^{-1}$. Using 
the older (pre-BATSE) fluence logN-logS relations, with 
relatively larger uncertainties, gives a factor of 2 smaller 
value: from Higdon and Lingenfelter (1990), we 
would derive  0.12 day$^{-1}$.
We adopt the (larger) BATSE value, which will impose 
a more conservative limit on beaming, and which is 
already comparable to the total observed rate of  AT 
sources given by PM: R$_{AT} \sim$ 0.3 day$^{-1}$.  
Given the \ga25\% uncertainties in R$_{X\gamma}$ and the AT source 
fluxes and exposure/rate estimates, which would yield 
comparable uncertainties in derived rates (given the local 
slope of logN-logS; see below), this may be viewed 
as satisfactory agreement. This would imply that the AT sample does 
not contain GRB afterglows that would not have been 
detected also as prompt GRBs. However, a closer examination 
of the AT data is warranted. 

\medskip

\noindent
{\it Contribution of X-ray Flares/Transients}

PM concluded that many (or most ?) of the AT sources 
were x-ray transients of known variety: Be-X-ray binaries, 
RS CVn stars, BL Lacs, and possibly known LMXBs. 
We also find (see below) that flare stars (dMe stars) 
are likely contributors. We summarize these possible 
identifications and thus the reduced sample of sources 
which could be either prompt or afterglow emission 
from GRBs. 

Four of the 27 AT sources are very possibly 
flaring outbursts from known Be binary systems: 
Gamma Cas, 4U0115+63, EXO2030+375, and 
3A2237+608=1H2214+589 since all were relatively long 
duration detections ($>$48, 36, $>$72 and $>$14 hours, respectively). 
An additional five sources are likely associated with RS CVn 
systems (HR1099, Sigma Gem, DM UMa, V532 Cen, 
and II Peg), one with a BL Lac (Mkn 421) 
and one with an LMXB (4U2127+119 in the globular cluster 
M15; or possibly a second LMXB and thus transient in this 
cluster, as we have in fact suggested previously (Grindlay 1993)) since 
again all were detected for more than several satellite 
orbits (6, 36, 5, 5, 3, 72 and $>$48 hours, respectively) and 
all had persistent {\it Ariel V} sources in the AT error boxes.  One additional 
AT source can also be rejected as unlikely to be a GRB (afterglow), 
given its long detection period (between 13 and 36 hours) and its 
possible association with a known (persistent) source, 
3A1956+034; one is probably associated with a symbiotic 
star and persistent source, 3A1703+241;
and one source (AT2030-330) is likely to 
be an active dMe star (AT Mic), as pointed out by 
Connors et al (1986). 

This leaves 13 AT sources as candidate GRB events. 
Their spatial distribution appears even more isotropic than the full 
AT sample, since several low latitude sources (e.g. the 4 Be 
transients) have been removed. 
This cleaned sample of AT-GRB candidates may be an underestimate, 
since two of these 13 sources are the historical GRBs (751102 and 760128) 
noted above and yet the error box for 
one of these (AT0318+220=GRB751102) also contains a persistent 
{\it Ariel V} source (3A0322+277) which is the RS CVn UX Ari. This likely 
false association with a RS CVn (we regard the temporal 
coincidence with a catalogued GRB as a certain identification) may imply  
that at least 1 out of 5 of the claimed RS CVn associations could be  
spurious. This might then imply that 1-2 
other identifications are accidental, and that the GRB candidates 
would thus be correspondingly increased. 
However this is balanced by the possibility of additional dMe star 
identifications for some, and we conclude that 13$\pm$3 sources and 
thus implied rate of AT-GRB candidates 
R$_{AG} \sim$ 0.15$\pm$0.04 day$^{-1}$
is a best estimate. Since this 
is (still) consistent with the expected rate of prompt 
GRB events, R(F\ga F$_{\gamma,min}) \sim$ 0.23 day$^{-1}$, 
it would again seem that only prompt GRBs are detected as AT sources. 
However the two known GRBs were not detected only for 
the minimum integration time of 1.6 hours (= 1 orbit), as 
expected if only prompt GRB emission is detected,  but rather for 
4.7 hours (GRB751102) and 3.1 hours (GRB760128). Thus x-ray 
afterglow detection, in addition to prompt GRB emission,  
is suggested  for these events and, from their durations (see below), 
for many or most of the others. 
Note also that prompt GRB emission with duration less 
than the 6sec {\it Ariel V} spin period was correspondingly less 
likely to be detected than afterglow emission from the same burst.

\section{Analysis of AT-GRB Candidates: Afterglows ?}

We now consider the 13 AT sources which are candidates for 
GRB afterglows detected at 2-10 keV. If these are afterglows, 
and not just prompt GRB (x-ray) emission, their light curves should 
exhibit roughly power-law (index \about1.3) decays and thus 
the brightest AT sources should have the longest detectable 
durations. However, for relatively weak GRBs detected 
near the end of a 6000s (1 AT orbit) integration time, they may  
be (still) detectable in the next orbit, so the correlation between peak 
flux and duration will be reduced. In fact for an afterglow with x-ray 
flux F$_x$ = F$_o$t$^{-1.3}$, an event 
occuring T seconds {\it before} the end of the first orbit (i.e. T \la 6000 s) 
will yield average flux measures F$_{x,1} \sim$ F$_o$/(0.3 T) 
for the first orbit and F$_{x,2} \sim$ F$_o$/(1800 T$^{0.3}$) 
for the next (second) orbit. Thus the flux ratio expected for 
an afterglow detected on two consecutive orbits is 
F$_{x,2}$/ F$_{x,1}$ = 6000/T$^{1.3}$ \ga 1 for T \la 811 sec (i.e. 
the mean flux would appear to rise rather than decay), whereas for a more 
probable T $\sim$ 3000 s, the ratio is 0.18 (falling to 0.07 
at the maximum T = 6000 s). Given the 
{\it Ariel V} AT source flux threshold of  S$_o$ = 20 mCrab, 
an average GRB arrival time (T = 3000 s; or midway through 
the first {\it Ariel V} orbit) and 
afterglow would then be detectable on the second orbit if 
it has a peak x-ray flux F$_o$ \ga 1800 S$_o$ T$^{0.3}$ \about 
2 $\times$ 10$^4$ S$_o$ so that its x-ray flux averaged over the 
first orbit must exceed F$_{x,1}$ \about (2 $\times$ 10$^4$/900) S$_o$ 
\about22 S$_o$. This in turn would suggest, for GRBs with logN-logS 
index \about 0.8 (see below), that 2-orbit afterglows would be detected 
only for \about(22)$^{-0.8}$ \about 1/12 of the sample, or \about 1 AT source. 
\bigskip

\small
\centerline{Table 1: Summary of AT-GRB Candidates}
\vskip 2pt

\begin{center}
\begin{tabular}{lrllll}
AT Source&  Yr-Mo-Da & Peak Flux (error) $^1$ & Duration$^2$ & LC Type$^3$ &  Comments\\ 
\hline &&&&& \\

AT1812+139 & 791206 & 9.1 (1.6) & 1.6 & 1 & marginal detection \\
AT0214+013 & 790626 & 9.8 (1.8) & 1.6 & 1 &   \\
AT0934+475 & 780826 & 10.6 (1.5) & 1.6 & 1 &   \\
AT1742-325 & 741124 & 14.5 (1.9) & 4.7 ? & 3 &  gap in LC coverage \\
AT0833-602 & 760414 & 13.2 (2.6) & 1.6 & 1,2 & possible 2 orb. detection  \\
AT1933-312 & 770831 & 17.4 (3.4) & 1.6 & 5 &   \\
AT1253-190 & 770328 & 18.0 (3.0) & 1.6 & 4,5 &  \\
AT2337+192 & 750819 & 19.3 (3.3) & 1.6 & 1 &  \\
AT1234+099 & 790823 & 22   (1.8) & 3.1 & 4 & pk. flux from LC, not table \\
AT1526+171 & 760618 & 25.8 (3.6) & 1.6 & 1,2 & possible 2 orb. detection \\
AT0318+220 & 751102 & 41.8 (8.1) & 4.7 & 2 & coincident with GRB751102 \\
AT0710+012 & 760128 & 53.3 (5.3) & 3.1 & 4 & coincident with GRB760128 \\
AT0323-284 & 771220 & 56.6 (4.9) & 1.6 & 1 &   \\
\end{tabular}
\end{center}
\vskip 5pt

\footnotesize
\noindent
Notes:\\
1. Peak flux in SSI counts/sec (from PM). \\
2. Duration (hours) is quantized in 1.6h (= 1 orbit) units (from PM).\\
3. Light Curve type (extracted from plots in  PM):\\
\hspace*{1cm} 1. single orbit detection; \\
\hspace*{1cm} 2. \ga2 orbit detection and possible smooth decay;\\
\hspace*{1cm} 3. \ga2 orbit detection but irregular;\\ 
\hspace*{1cm} 4. possible precursor emission;\\
\hspace*{1cm} 5. gap in LC after peak. 
\medskip

\normalsize

In Table 1 we list the 13 AT-GRB 
candidates ordered by peak flux as given in PM.
There is marginal evidence that the brighter AT events are longer 
duration, particularly if the gaps in the coverage (LC type 5 in table) 
are considered. However, there are some notable exceptions: the brightest 
event (771220) is detected in only 1 orbit, and the much fainter event 
(741124) is detected over at least 3 orbits. The brightest event could 
be explained as a prompt GRB detected through the detector body but yet  
outside the collimator field of view so that only down-scattered hard x-rays 
or $\gamma$-rays from the GRB are recorded, and the fainter event could 
be a (non-rejected) background transient/flare source. Of the 13 AT-GRB 
candidates, only 5 are constrained to be \la1.6h (single orbit) 
detections, and 3 of these are the faintest sources nearest the detection limit. 
Although 4 additional (9 total) events are listed (Table 1) as 1.6h duration, 
their light curves or coverage gaps do not require them to be. 
Thus \about 4-8 of the AT-GRB candidates are consistent with 
\ga2 orbit AT source detections and thus afterglow emission, or (well) 
in excess of that estimated for a ``mean" GRB afterglow. Given the 
simplfifying assumptions (uniform power law decay) and small N 
sample (and thus sampling of the integration time T within 
the first orbit), we conclude that the entire 13 source sample provides a 
conservative  upper limit to the rate of GRBs detected only 
by their afterglow: R$_{AG}$(F$_{\gamma,min}$ \ga 1.2 $\times 10^{-5}$ \ergcm2)  
$\sim$ 0.15 day$^{-1}$. 

\section{Other Fast Transient Surveys and Constraints}

In addition to the AT survey and the several earlier results on fast 
transients (FTs) referenced in PM, the HEAO-A1 (Ambruster and Wood 1986) 
and A2 instruments (Connors, Serlemitsos and Swank 1986) provided 
additional constraints on the population of FTs and their possible 
relation to GRBs. Both studies recognized the probable inclusion of 
prompt emission from GRBs but did not, of course, consider afterglows.

\medskip

\noindent
{\it HEAO-A1 Survey}

Ambruster and Wood (1986=AW) reported the analysis of the first 6
months of HEAO-1 scanning data from the A1 detectors. Ten FTs were 
detected above a flux threshold of f$_{A1}$(2-10 keV) = 
7 $\times 10^{-11}$ \erg/cm2sec. The A1 detectors used had field of 
view 1\deg $\times$ 4\deg (FWHM) and scanned the sky with 35min 
scan period for a $\sim$10s exposure time/scan on a given source. 
Thus for FTs defined (conservatively) as having duration \la1 scan 
period, the threshold x-ray fluence detected was 
F$_{A1,min} \sim$ (60)(35) f$_{A1}$ =  1.5 $\times 10^{-7}$ \ergcm2. 
As for the analysis of the AT events above, we let 
F$_{X,min}$ = F$_{A1,min}$ and then scale by the x-ray/$\gamma$-ray 
fluence ratio, r$_{X\gamma,SAX}$, to derive  
F$_{\gamma,min}$ = F$_{X,min}$/r$_{X\gamma,SAX}   
\sim 7.5 \times 10^{-7}$ \ergcm2 as being the corresponding
(approximate) fluence threshold for GRBs in the BATSE band. 
 
AW derived an all-sky rate for their FT sample of $\sim$1500
yr$^{-1}$. Since their sample contained 6 (out of 10) ``identified" 
stellar sources (RS CVns, dMe stars) or catalogued A1 sources, we 
regard only the remaining four (H0422+37, 0535-86, 1908-18, 
and 0547-14) as GRB candidates. Indeed, two of these (1908-18 
and 0547-14) are already listed as GRB candidates based on their 
relatively hard spectra (though no simultaneous GRB was 
reported by other experiments at the time). The resulting point 
on a logN-logS (fluence) plot is then 
N(F$_{\gamma,min}$  \ga7.5 $\times 10^{-7}$) = 600yr$^{-1}$ =
1.6day$^{-1}$, with errors of at least a factor of two due to the 
small number statistics (and flux conversion uncertainties). 
Still, it is interesting that this logN-logS point is in fact 
nicely consistent with the BATSE curve of  Bloom et al (1996), allowing 
for BATSE coverage (\about0.4) 
 and efficiency ($\sim$0.6) at such low fluences, 
which would predict a total GRB rate of $\sim$2day$^{-1}$. When ``connected" 
to the logN-logS point derived above for the AT sample, the A1-AT 
slope would be \about0.8 (again, consistent with the 
approximate BATSE slope at F$_{\gamma,min}$). 
Thus the A1 sample provides independent confirmation of the AT 
result: x-ray detection of FTs (without known GRBs) yields 
a detection rate and logN-logS consistent with the known BATSE 
GRB rate. No additional excess rate is implied, as might arise 
if the x-ray afterglow emission were less beamed than the prompt 
GRB emission.
\medskip

\noindent
{\it HEAO-A2 Survey}

The A2 results for FTs (Connors et al 1986) yielded rather 
similar results: 8 FTs were detected in 3 survey passes (over 
1.5 yr)  of the 3 A2/HED detectors scanning (35 min period) through the 
sky. At least 3 FTs are directly attributable to stellar flaring 
sources (AT Mic and other dMe stars, etc.). The quoted flux limit 
of 4 mCrab is also similar, and thus the logN-logS constraint for 
the \la5 GRB candidates would also be similar or even lower in rate 
(given the \ga3$\times$ exposure). 
Assuming none of the events are from GRBs (but all are stellar
flares), the authors derive an upper limit on the GRB rate of 
1300 yr$^{-1}$, or within a factor of $\sim$2 of the A1 rate.

\section{Discussion}
The approximate (factor of \about2) 
agreement in rates, and derived logN-logS, for 
GRBs possibly detected by their x-ray afterglow 
as FTs and that expected from 
current BATSE logN-logS limits rules out strong beaming
differences (e.g. factors of \ga3-10)  
between prompt GRB ($\gamma$-ray) emission and 
x-ray afterglows. Although all three x-ray surveys for FTs 
are significantly contaminated (perhaps dominated) by stellar 
flare sources, they all contain similar fractions ($\sim$0.5) 
of events that are consistent with being GRB emission. The AT 
sample provides the strongest evidence for these, with two 
confirmed detections (simultaneous GRBs) and yet FT durations 
longer than 1.6h which strongly suggest that x-ray 
afterglow emission was detected. 

Although differential beaming  between afterglow vs. prompt 
emission is not required, factors of \about3 would still 
be allowed by these results.  
More sensitive x-ray (and hard x-ray) surveys and monitors could 
provide much more restrictive limits. The present limits are 
set primarily by the limited duty cycle of the sky coverage for 
a given source (GRB) imposed by the narrow-field scanning 
instruments on {\it Ariel V} or HEAO-1. The conservative flux 
limits presented here were derived assuming the source was 
present for a full integration time or 
scan (1.6h or 35min, respectively). With 
a wide-field imaging instrument providing both much larger 
duty cycle coverage as well as imaging capability to reject 
foreground stellar and background AGN flare sources, these limits could 
be improved significantly. It is interesting 
that 1970s-era x-ray (2-10 keV) 
missions and sensitivities were, in retrospect, able to 
provide rough constraints for GRB 
logN-logS values at levels approaching current BATSE limits 
(i.e. at approximately F$_{\gamma,min}$ \ga 
(0.075 - 1.2) $\times 10^{-5}$ \ergcm2). 

In one sense the limits presented here are strict lower limits 
on the need for differential beaming: they assume that every GRB 
is accompanied by afterglow x-ray emission and that the
x-ray/$\gamma$-ray fluence ratio is constant at the Ginga/SAX value 
of $\sim$0.2. Neither is strictly true (e.g. x-ray afterglows 
have thus far been reported by BeppoSAX only for the ``long'' 
GRBs), and reductions (most likely) 
in either parameter will increase the effective rate of x-ray afterglow 
GRBs to larger values by a reciprocal factor. On the other hand, 
the  limits imposed by fast transients 
are upper limits due to residual contamination by 
flaring x-ray sources (dMe stars, etc.). Planned 
all-sky x-ray (2-10 keV) monitors with \about1-3  
mCrab sensitivity and coarse imaging capability 
(e.g. MOXE on Spectrum-X [in't Zand et al 1994] 
or MAXI for JEM on ISS [Kawai et al 1996]), as well as 
proposed hard x-ray imaging surveys (e.g. EXIST 
[Grindlay et al 1995]) and possible rapid-slew 
narrow-field x-ray detectors should allow 
the prompt vs. afterglow relative beaming factor as 
well as x-ray/$\gamma$-ray fluence ratios to be derived.

\bigskip

I thank V. Petrosian and J. Bloom for comments on logN-logS relations 
for burst fluence and A. Connors for comments on the HEAO-A2 data.  
This work was supported in part by NASA grants NAG5-3256 and NAG5-3808.

\end{document}